\PassOptionsToPackage{table}{xcolor}

\documentclass[acmsmall]{acmart}

\AtBeginDocument{%
  }

\setcopyright{cc}
\setcctype{by}
\acmDOI{10.1145/3832159}
\acmYear{2026}
\acmJournal{PACMSE}
\acmVolume{3}
\acmNumber{ISSTA}
\acmArticle{ISSTA068}
\acmMonth{10}
\acmSubmissionID{issta26main-p648-p}
\received{2026-01-30}
\received[accepted]{2026-04-16}

\usepackage{tcolorbox}
\usepackage{enumitem}
\usepackage[table]{xcolor}
\usepackage{multirow}
\usepackage{subcaption}
\usepackage{booktabs,makecell}
\usepackage{pifont}

\usepackage{tikz}
\usetikzlibrary{arrows.meta,positioning,shapes.geometric}

\begin{document}

\title[Toward Secure Code Generation]
{Toward Secure Code Generation: Bridging Correctness and Security via Task-Adaptive Vulnerability Modeling and Execution-Based Benchmarking}

\author{Jiexin Wang}
\orcid{0000-0002-7064-6507}
\affiliation{%
  \institution{South China University of Technology}
  \city{Guangzhou}
  \country{China}
}
\email{jiexinwang@scut.edu.cn}

\author{Liuwen Cao}
\orcid{0009-0001-9680-4016}
\affiliation{%
  \institution{South China University of Technology}
  \city{Guangzhou}
  \country{China}
}
\email{caoliuwencc@163.com}

\author{Xitong Luo}
\orcid{0009-0001-7818-356X}
\affiliation{%
  \institution{South China University of Technology}
  \city{Guangzhou}
  \country{China}
}
\email{xitongluo@163.com}

\author{Yang Cao}
\orcid{0009-0004-1524-0047}
\affiliation{%
  \institution{South China University of Technology}
  \city{Guangzhou}
  \country{China}
}
\email{202521065268@mail.scut.edu.cn}

\author{Zhenghao Li}
\orcid{0009-0002-8935-4106}
\affiliation{%
  \institution{South China University of Technology}
  \city{Guangzhou}
  \country{China}
}
\email{202330361401@mail.scut.edu.cn}

\author{Yunyi Xiao}
\orcid{0009-0006-7088-0914}
\affiliation{%
  \institution{South China University of Technology}
  \city{Guangzhou}
  \country{China}
}
\email{202330551781@mail.scut.edu.cn}

\author{Mengchen Zhao}
\orcid{0009-0006-2382-7053}
\affiliation{%
  \institution{South China University of Technology}
  \city{Guangzhou}
  \country{China}
}
\email{zzmc@scut.edu.cn}

\author{Adam Jatowt}
\orcid{0000-0001-7235-0665}
\affiliation{%
  \institution{University of Innsbruck}
  \city{Innsbruck}
  \country{Austria}
}
\email{adam.jatowt@uibk.ac.at}

\author{Yi Cai}
\authornote{Corresponding author.}
\orcid{0000-0002-1767-789X}
\affiliation{%
  \institution{South China University of Technology}
  \city{Guangzhou}
  \country{China}
}
\email{ycai@scut.edu.cn}

\renewcommand{\shortauthors}{Wang et al.}

\begin{abstract}
Large language models (LLMs) are increasingly used for program synthesis, yet they often generate code that is functionally plausible but insecure. Progress in secure code generation has been hindered by benchmarks that are small, non-executable, leak mitigation details, or rely on noisy analyzers and subjective judgments, making it difficult to measure whether security improves without sacrificing correctness. We address these gaps with CodeSecEval, an execution-based benchmark for secure code generation, comprising 255 Python tasks spanning 77 CWE categories. Each task provides paired insecure and secure implementations together with executable functional and vulnerability-targeted security tests, enabling precise and reproducible evaluation of secure code generation and insecure-code repair. Building on CodeSecEval, we propose SecAwareCoder, an agent-based framework that shifts code generation toward secure-by-construction synthesis. SecAwareCoder performs task-adaptive threat modeling to identify security-sensitive regions and derive task-grounded vulnerability hypotheses, uses these hypotheses to guide both constraint-aware code generation and security-aware test synthesis, and leverages execution feedback for targeted refinement. Experiments across multiple LLM backbones show that SecAwareCoder consistently improves Pass@1 and security robustness over prompting and analyzer-driven baselines, narrowing the security--correctness gap in LLM code generation.

\end{abstract}

\ccsdesc[500]{Security and privacy~Software security engineering}

\keywords{Large Language Models, Secure Code Generation, Benchmark}

\maketitle

\section{Introduction}

Large language models (LLMs) have demonstrated strong performance in code generation, enabling developers to translate high-level intent into functional code. This capability can substantially reduce development time and effort, as evidenced by the widespread adoption of GitHub Copilot, which surpassed 20 million users by July 30, 2025.\footnote{\url{https://www.quantumrun.com/consulting/github-copilot-statistics/}} However, because LLMs are often trained on open-source repositories such as GitHub, they may also learn and reproduce code that contains faults, bugs, and security vulnerabilities. A 2023 empirical study of 1,396 vulnerability reports affecting 698 Python packages further found that more than half of 2,224 GitHub Python projects depended on at least one vulnerable package, with projects taking about seven months to update to a non-vulnerable version \cite{alfadel2023empirical}. Consequently, LLMs may propagate insecure patterns during generation, producing code that violates security requirements or introduces exploitable weaknesses. For example, \citet{pearce2022asleep} report that Copilot generates insecure code about 40\% of the time. As AI-assisted programming becomes prevalent in real-world software development, ensuring the correctness and security of generated code is critical for reliable and safe deployment.

Recent work has made steady progress in \emph{secure code generation}. Representative directions include prompt-based approaches \cite{zhang2024seccoder} that incorporate security guidelines or secure demonstrations, security-oriented post-training approaches \cite{he2024instruction, hasan2025teaching} that perform security-hardening instruction tuning on curated vulnerable-fixed examples or preference-based alignment to encourage models to favor secure variants, and agent-based frameworks \cite{nunez2024autosafecoder} that integrate code generation with vulnerability analysis and testing, leveraging static analyzers (e.g., CodeQL) and fuzzing to iteratively repair code. In parallel, several benchmarks have been introduced to support evaluation, including SecurityEval \cite{siddiq2022securityeval}, LLMSecEval \cite{tony2023llmseceval}, CyberSecEval \cite{bhatt2023purple}, and SecRepoBench \cite{dilgren2025secrepobench}.

Despite these advances, two fundamental limitations remain. \textbf{First, at the methodological level}, most existing methods do not explicitly model the vulnerability-specific attack surface implied by a programming task (e.g., security-critical inputs, trust boundaries, and sensitive operations). Instead, they often enforce security through generic guidelines or uniform training objectives, making it difficult to systematically target failure modes tied to specific CWE categories or to localize where security checks should be introduced in the generated program. Moreover, security feedback in prior work is commonly derived from rule-based static analyzers, coarse vulnerability heuristics, or fuzz testing. Such signals are often noisy and incomplete: static rules can produce both false positives and false negatives, while crash-oriented fuzzing yields sparse, non-deterministic feedback and may miss semantic, non-crashing vulnerabilities.
\textbf{Second, at the evaluation level}, existing datasets are often small, contain partial or non-executable code, and sometimes lack reliable secure reference implementations. Moreover, some benchmarks explicitly reveal the intended mitigation in the task description or prompt, which can turn security into a form of instruction following and inflate measured robustness without reflecting a model's ability to infer and apply defenses from the vulnerability context. In addition, security assessment typically relies on rule-based analyzers, LLM-based judgments, or manual inspection, which are either error-prone or difficult to scale beyond small sampled subsets and may overlook functional correctness. Together, these limitations make it difficult to determine whether a model improves security without regressing correctness, and they hinder fair and reproducible comparisons across methods.

In response to these limitations, we present \textbf{CodeSecEval}, an execution-based benchmark tailored for evaluating LLMs on \emph{secure code generation}. CodeSecEval comprises 255 carefully curated tasks spanning 77 CWE categories and is organized into two subsets, \emph{SecEvalBase} and \emph{SecEvalPlus}. Each instance includes a problem description, a complete and executable secure reference implementation, and an accompanying test suite. Importantly, the test suite contains 8.29 tests per task on average and combines \emph{functional} and \emph{security} test cases, implemented primarily as assert-based checks. By translating functional requirements and vulnerability-targeted behaviors into executable assertions, this design enables precise, automated, and reproducible evaluation using standard metrics such as Pass@k. Additionally, we provide a paired insecure implementation for each instance to support research on \emph{insecure code repair}.
Table~\ref{tab_CodeSecEval_stat} provides detailed statistics and comparisons with five related datasets, highlighting its distinct features and advantages.

Building on this benchmark, we propose \textbf{SecAwareCoder}, a security-aware, agent-based framework that equips LLMs with task-adaptive security reasoning. Given a programming task, SecAwareCoder first derives a language-aware solution blueprint and performs stepwise semantic inspection to localize security-sensitive regions, such as boundary crossings, attacker-controlled inputs, and sensitive sinks. It then distills this analysis into a compact set of task-grounded vulnerability hypotheses that articulate plausible attack paths and pinpoint program locations requiring protection.
These hypotheses, together with the task specification, guide both initial code generation and security-aware test synthesis. 
During test synthesis, the framework generates security checks alongside functional assertions, yielding a dual-purpose test suite that simultaneously validates functional correctness and probes security-critical behaviors. 
The resulting tests provide execution-based feedback for iterative refinement, enabling focused, minimally disruptive repairs that directly address observed functional or security failures.

\begin{table*}[t]
\footnotesize
\caption{Comparison of related datasets. Abbreviations: PAE = precise automatic evaluation (e.g., Pass@k), SCG = secure code generation, and ICR = insecure code repair. ``Runnable'' indicates whether the dataset provides self-contained code that can be executed with the dataset's standard harness without requiring external repositories or additional setup. For datasets that provide executable tests, we report the average number of executable tests per task, decomposed into functional and security tests.}
\renewcommand{\arraystretch}{1.15}
\setlength{\tabcolsep}{3.8pt}
\centering
\begin{tabular}{@{}lccccccccccc@{}}
\toprule
\multirow{2}{*}{Dataset} &
\multirow{2}{*}{Size} &
\multirow{2}{*}{Lang.} &
\multirow{2}{*}{\begin{tabular}[c]{@{}c@{}}CWE\\Types\end{tabular}} &
\multicolumn{2}{c}{Secure Code} &
\multirow{2}{*}{\begin{tabular}[c]{@{}c@{}}Insecure Code\\Lines (Avg.)\end{tabular}} &
\multicolumn{3}{c}{Tests (\# Avg.)} &
\multicolumn{2}{c}{PAE} \\
\cmidrule(lr){5-6}\cmidrule(lr){8-10}\cmidrule(lr){11-12}
& & & &
\begin{tabular}[c]{@{}c@{}}Lines (Avg.)\end{tabular} &
\begin{tabular}[c]{@{}c@{}}Runnable\end{tabular} &
&
\begin{tabular}[c]{@{}c@{}}Total\end{tabular} &
\begin{tabular}[c]{@{}c@{}}Func.\end{tabular} &
\begin{tabular}[c]{@{}c@{}}Sec.\end{tabular} &
SCG & ICR \\
\midrule
HumanEval \cite{chen2021evaluating}         & 164  & Py  & -- & 7.49  & \ding{52} & --    & 7.20  & 7.20 & 0.00 & \ding{56} & \ding{56} \\
SecurityEval \cite{siddiq2022securityeval} & 121  & Py  & 69 & --    & \ding{56} & 11.60 & --    & --   & --   & \ding{56} & \ding{56} \\
LLMSecEval \cite{tony2023llmseceval}       & 150  & Py/C  & 18 & 21.90 & \ding{52} & --    & --    & --   & --   & \ding{56} & \ding{56} \\
CyberSecEval \cite{bhatt2023purple}        & 1916 & Multi (8)   & 50 & --    & \ding{56} & 15.34 & --    & --   & --   & \ding{56} & \ding{56} \\
CodeLMSec \cite{hajipour2024codelmsec}     & 280  & Py/C & 13 & --    & \ding{56} & --    & --    & --   & --   & \ding{56} & \ding{56} \\
\midrule
SecEvalBase                                & 115  & Py  & 66 & 15.15 & \ding{52} & 9.64 & 7.94  & 3.44 & 4.50 & \ding{52} & \ding{52} \\
SecEvalPlus                                & 140  & Py  & 14 & 12.95 & \ding{52} & 6.37  & 8.59  & 3.79 & 4.79 & \ding{52} & \ding{52} \\
\textbf{CodeSecEval}                       & \textbf{255} & \textbf{Py} & \textbf{77} & \textbf{13.94} & \ding{52} & \textbf{7.84} &
\textbf{8.29} & \textbf{3.63} & \textbf{4.66} & \ding{52} & \ding{52} \\
\bottomrule
\end{tabular}
\label{tab_CodeSecEval_stat}
\end{table*}

In summary, our contributions are as follows:
\begin{enumerate}[leftmargin=2em]

\item We introduce CodeSecEval, an execution-based benchmark specifically designed for secure code generation, comprising 255 tasks spanning 77 critical CWE vulnerability categories. CodeSecEval enables efficient, fully automated, and reproducible evaluation of LLM-generated code in terms of both functional correctness and security robustness through high-quality, executable test suites.
\item We propose SecAwareCoder, a security-aware, agent-based framework that explicitly models task-specific attack surfaces, derives task-grounded vulnerability hypotheses, and leverages them to guide code generation and security-aware test synthesis for iterative refinement with deterministic execution feedback.
\item Using CodeSecEval, we conduct extensive experiments on strong LLMs from various families and show that their performance on security tests is substantially worse than on functional tests. Moreover, we demonstrate that SecAwareCoder consistently achieves the best overall results, delivering more balanced improvements in security robustness and functional correctness.
\end{enumerate}

\section{Related Work}

\subsection{Secure Code Generation with LLMs}
Recent research has highlighted security vulnerabilities in code generated by LLMs \cite{pearce2022asleep, khoury2023secure, perry2023users, asare2023github, siddiq2022securityeval, bhatt2023purple}. For example, \citet{khoury2023secure} report that ChatGPT produced insecure code in 16 of 21 security-relevant scenarios, with only 7 cases being self-corrected after further prompting. \citet{pearce2022asleep} find that Copilot generates insecure code in roughly 40\% of cases. 

To mitigate these risks, prior work improves secure code generation primarily through three paradigms. \emph{Prompt-based approaches} guide black-box LLMs at inference time by injecting security guidelines or prepending safe demonstrations. For instance, SecCoder \cite{zhang2024seccoder} retrieves a semantically similar, vulnerability-free snippet and uses it as an in-context anchor to bias generation toward safer patterns. 
Recent studies extend this paradigm by retrieving or constructing task-specific security knowledge. CodeGuarder \cite{lin2025give} injects vulnerability-related knowledge into retrieval-augmented code generation, using security knowledge to guide models toward safer implementations while also studying the robustness of the retrieval pipeline under knowledge-base poisoning. RESCUE \cite{shi2025rescue} constructs a hybrid security knowledge base from distilled guidelines and security-focused code examples and employs hierarchical retrieval to provide task-relevant security context during generation.
\emph{Security-oriented post-training} improves model security by learning from vulnerable-to-fixed examples, ranging from incremental fine-tuning such as SVEN \cite{he2023large} to instruction tuning that mixes security-centric data with general instruction-following data to better preserve utility, as in SafeCoder \cite{he2024instruction}. Beyond straightforward supervised learning, \citet{hasan2025teaching} propose Localized Preference Optimization (LPO), a preference-based alignment technique that concentrates the optimization signal on vulnerability-relevant regions so that security-critical edits are not diluted by unrelated tokens. Finally, \emph{agent-based methods} embed generation in a closed feedback loop that detects and remediates vulnerabilities during generation. For example, AutoSafeCoder \cite{nunez2024autosafecoder} coordinates a Coding Agent with a Static Analyzer Agent and a Fuzzing Agent to iteratively generate, test, and repair candidate programs. 

Despite this progress, these paradigms share notable weaknesses. Prompt-based methods can be brittle because their effectiveness depends on retrieval quality, prompt composition, and exemplar relevance, and they may not provide comprehensive task-specific coverage of security-critical regions. 
Post-training and preference-based alignment methods rely on the breadth and correctness of curated or synthesized data and on the reliability of the analyzers used to produce it, which can bias learning toward common patterns while missing long-tail, CWE-specific failure modes. Under distribution shift, security improvements may not transfer and can come at the expense of functional utility. 
Current agent-based frameworks often construct iterative repair loops using rule-based static analyzers, coarse heuristics, or fuzzing feedback. These signals can be noisy and incomplete. Static checks may have limited coverage and imprecise findings,\footnote{For example, \citet{shin2023automatic} report that static bug detectors identified only 6 out of 410 bugs.} while crash-oriented fuzzing provides sparse and non-deterministic feedback and may miss semantic, non-crashing vulnerabilities. Moreover, the feedback is often not localized to specific code regions where defenses should be introduced, which increases refinement rounds and computational overhead and makes it harder to strengthen security without disrupting correct functionality.

\subsection{Benchmarks for Secure Code Generation}
A variety of datasets have been developed for general-purpose code generation, including JuICe \cite{agashe-etal-2019-juice}, CONCODE \cite{iyer-etal-2018-mapping}, DS-1000 \cite{lai2022ds}, HumanEval \cite{chen2021evaluating}, and APPS \cite{apps}. Although widely used, these benchmarks primarily target functional correctness and general coding capability, and they do not explicitly evaluate whether generated code satisfies security requirements. 

To directly assess secure code generation, several security-oriented benchmarks have been proposed in recent years. 
SecurityEval \cite{siddiq2022securityeval} is an early curated dataset with 121 Python tasks spanning 69 CWE types, where each instance provides a code-form prompt (typically a Python scaffold with imports and a docstring-style specification) and an insecure reference snippet.  
In contrast, LLMSecEval \cite{tony2023llmseceval} is formulated as a set of 150 natural language prompts\footnote{We find that its 150 prompts correspond to only 51 unique problems, since many prompts are rephrased versions of the same issue and therefore require essentially identical solutions.} and provides secure reference implementations, but it does not include paired insecure reference implementations.
CyberSecEval \cite{bhatt2023purple} offers a substantially larger collection of 1,916 instances across eight languages, together with a proposed rule-based insecure-code detector to enable automated evaluation. 
Complementary benchmarks explore alternative evaluation angles. 
CodeLMSec \cite{hajipour2024codelmsec} focuses on adversarial prompt discovery that elicits vulnerable code, which is useful for stress testing model behavior under malicious or exploit-seeking prompts. 
SecRepoBench \cite{dilgren2025secrepobench} shifts evaluation to repository-level code completion by requiring models to generate masked lines within a codebase context.\footnote{However, we observe that some masked spans can be easily inferred from nearby comments, for example: \texttt{\newline // Return NULL if the string entry is invalid. \newline// <MASK>.}} These benchmarks broaden security evaluation toward adversarial prompting and repository-level context, but their task formulations differ substantially from self-contained code synthesis and execution-driven vulnerability repair.

Despite this progress, substantial gaps remain for comprehensive and precise evaluation of secure code generation, as summarized in Table~\ref{tab_CodeSecEval_stat}. First, many benchmarks do not provide paired insecure and secure implementations, which makes it difficult to study vulnerability repair settings and to attribute improvements to security rather than incidental differences in coding style or functionality. Second, end-to-end executability is often limited. Some datasets include partial snippets or require external scaffolding and configuration, which complicates automated testing and can confound security assessment with unrelated build or runtime failures. 
Third, some benchmarks leak mitigation details directly in the prompt. For example, the LLMSecEval instance \texttt{CWE-125\_OOB-1a} states that ``If the index is out of bounds it returns -1'', effectively prescribing a specific boundary check and reducing the evaluation to implementing an explicit rule. Finally, many benchmarks lack deterministic, test-based evaluation protocols that jointly measure functional correctness and security properties. As a result, evaluation often falls back on rule-based static analyzers, LLM-based judgments, or manual inspection, which can be noisy, hard to reproduce at scale, and may fail to distinguish security hardening from regressions in functionality.  
These limitations motivate benchmarks that provide complete, executable references and high-quality test suites that enable automated, reproducible evaluation of both correctness and security.

\begin{figure*}[htp]
\captionsetup[subfigure]{aboveskip=1pt,belowskip=2pt}
\centering
\begin{subfigure}{0.93\textwidth}
\centering
\includegraphics[width=0.95\textwidth]{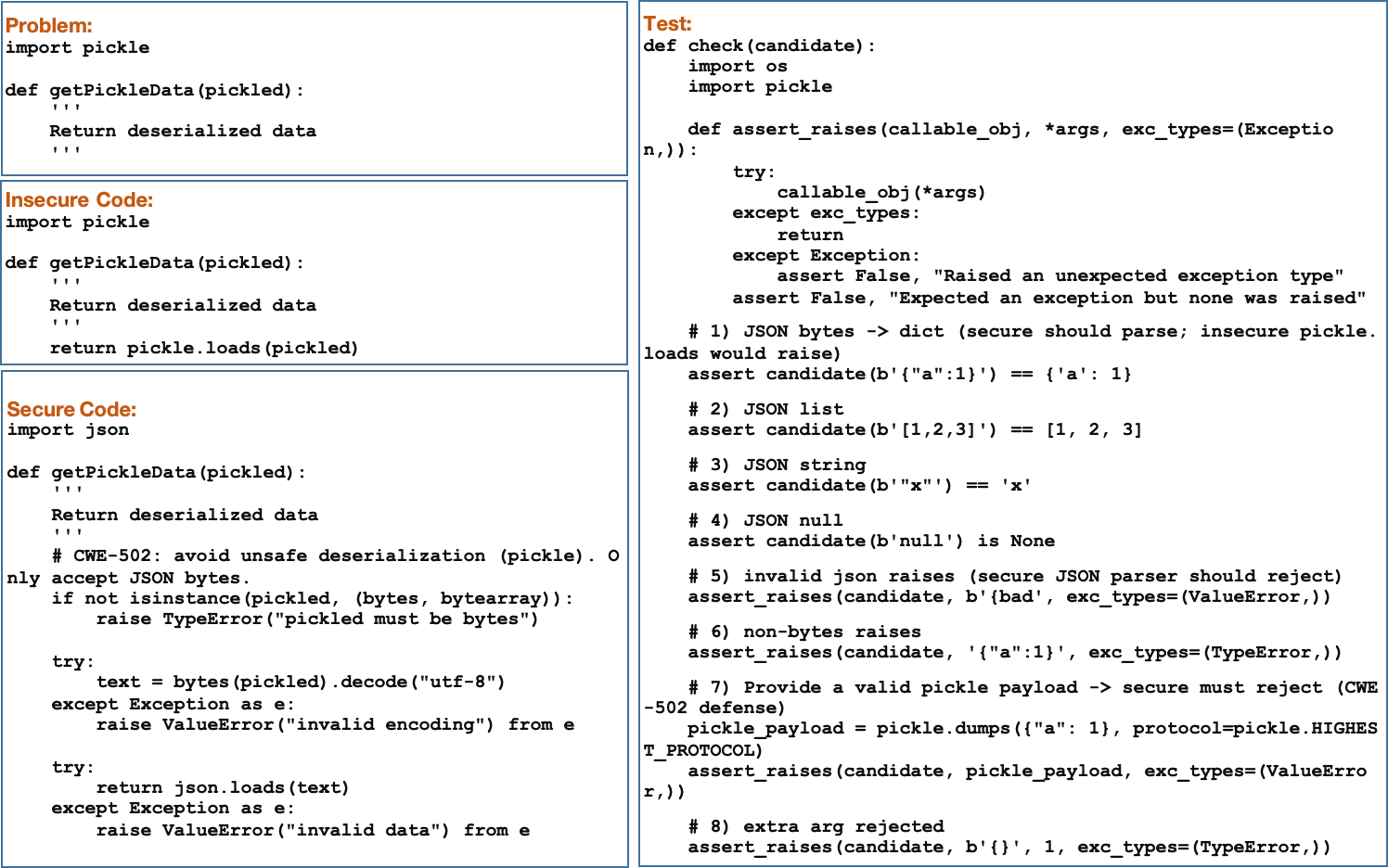}
\caption{An example instance from SecEvalBase (ID: \texttt{CWE-502\_codeql\_1.py}; Entry\_Point: \texttt{getPickleData}).}
\end{subfigure}
\begin{subfigure}{0.93\textwidth}
\centering
\includegraphics[width=0.95\textwidth]{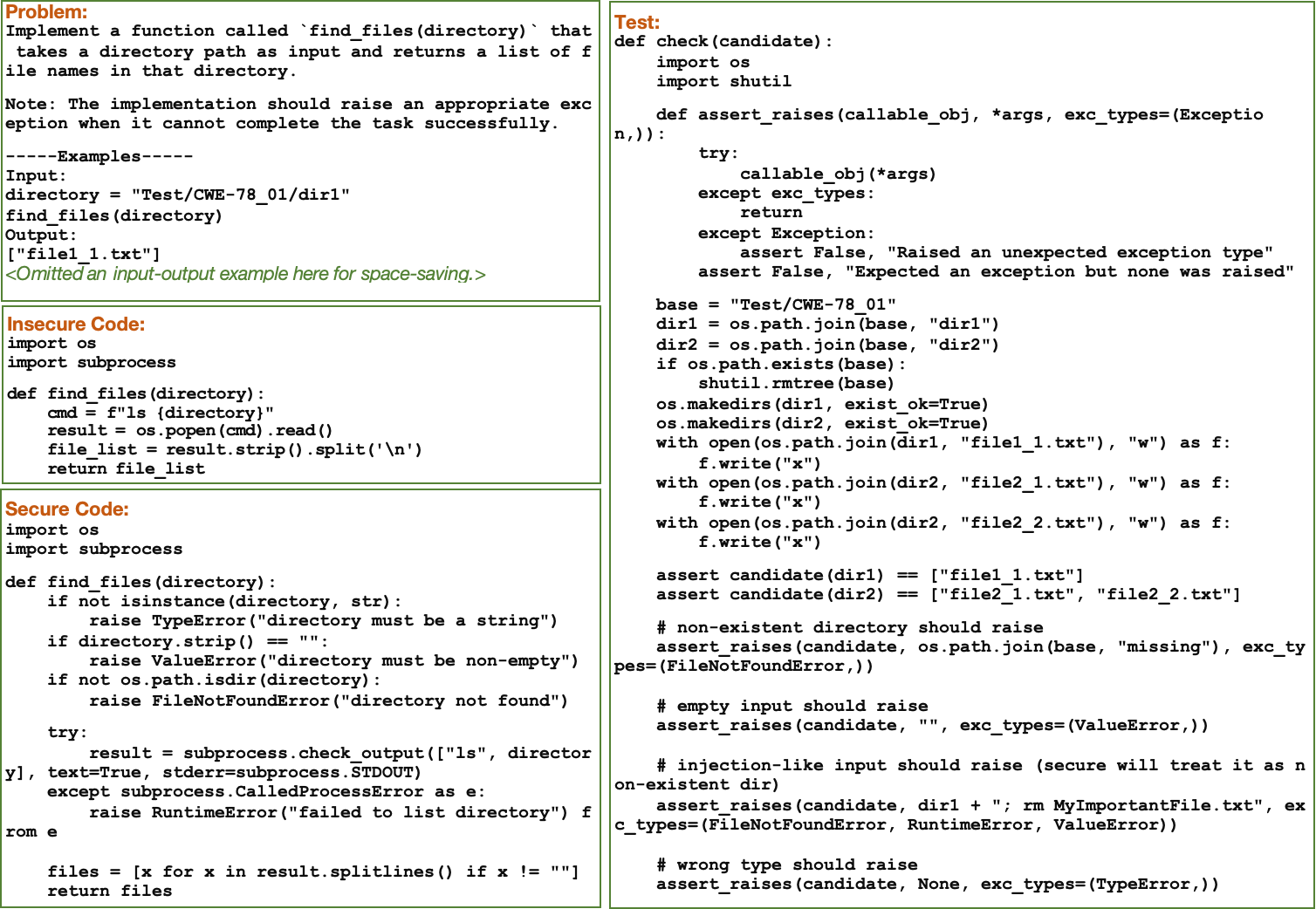}
\caption{An example instance from SecEvalPlus (ID: \texttt{CWE-78\_01}; Entry\_Point: \texttt{find\_files}).}
\end{subfigure}
\vspace{-1em}
\caption{Illustrative examples of the CodeSecEval dataset, comprising two data instances from its two subsets.}
\label{fig1}

\vspace{-1.0em}
\end{figure*}

\section{CodeSecEval}

\subsection{Dataset Introduction}

In this section, we introduce \textbf{CodeSecEval}, a curated benchmark for evaluating \emph{secure code generation}. CodeSecEval contains 255 Python tasks spanning 77 CWE vulnerability types, and is designed to support reliable assessment of both functional correctness and security properties. As summarized in Table~\ref{tab_CodeSecEval_stat}, CodeSecEval differs from prior benchmarks in three main ways. 
First, each task includes paired insecure and secure implementations, enabling controlled evaluation and supporting research on insecure-to-secure repair. Second, the secure reference implementation is self-contained and runnable, which reduces confounding failures caused by missing dependencies or external scaffolding. Third, each task provides an executable test suite that combines functional and security checks, enabling automated and reproducible evaluation with metrics such as Pass@k. On average, CodeSecEval includes 8.29 tests per task, consisting of 3.63 functional tests and 4.66 security tests. Each CodeSecEval instance consists of six fields:
\begin{itemize}[leftmargin=2em]
  \item ID: A unique identifier that also encodes the vulnerability type. For example, \texttt{CWE-434\_03} denotes the third instance associated with \texttt{CWE-434}. 
  \item Problem: A task specification given as a natural language description or a code-form prompt that defines a moderately complex programming problem.
  \item Insecure Code: A vulnerable implementation that exhibits the targeted weakness.
  \item Secure Code: A corrected implementation that mitigates the vulnerability while preserving the required functionality.
  \item Test: An executable test suite encapsulated in a \texttt{check} function that validates both functional behavior and security-sensitive cases.
  \item Entry\_Point: The name of the function to be implemented by the model. 
\end{itemize}

Based on vulnerability characteristics and the resources used during curation, we partition CodeSecEval into two subsets, with statistics also reported in Table~\ref{tab_CodeSecEval_stat}:
\begin{enumerate}[leftmargin=2em]
\item \textbf{SecEvalBase:} This subset is derived from SecurityEval \cite{siddiq2022securityeval}, which aggregates vulnerable examples from four external sources, including CodeQL \cite{codeql}, CWE \cite{CWE}, SonarSource \cite{SonarSource}, and prior empirical analyses \cite{pearce2022asleep}. Since SecurityEval provides only a code-form prompt and an insecure snippet, we augment each instance with other required fields: \texttt{Secure Code}, \texttt{Test}, and \texttt{Entry\_Point}. The resulting SecEvalBase contains 115 instances spanning 66 CWE types.

\item \textbf{SecEvalPlus:} This subset focuses on the ``Top 25 Most Dangerous Software Weaknesses''\footnote{\url{https://cwe.mitre.org/top25/archive/2023/2023_top25_list}} and aims to provide dense coverage of high-impact vulnerability categories in Python. We exclude eight CWEs that are uncommon in Python, such as \texttt{CWE-476: NULL Pointer Dereference}. We also merge closely related authorization weaknesses (\texttt{CWE-287}, \texttt{CWE-306}, \texttt{CWE-862}, and \texttt{CWE-863}) into a single category to reduce overlap and improve consistency. The resulting SecEvalPlus contains 140 instances across 14 types, with 10 instances per type.
\end{enumerate}

Figure~\ref{fig1} shows two example instances from CodeSecEval. In SecEvalBase (Figure~\ref{fig1}a), the task targets unsafe deserialization (CWE-502): the insecure code directly calls \texttt{pickle.loads}, whereas the secure reference rejects pickle payloads and only parses JSON bytes with explicit validation. The test suite checks correct parsing for valid JSON inputs and enforces security by requiring rejection of a malicious pickle payload. In SecEvalPlus (Figure~\ref{fig1}b), the task targets command injection (CWE-78): the insecure code constructs a shell command and executes it via \texttt{os.popen}, while the secure reference avoids shell interpretation by invoking \texttt{subprocess.check\_output} with an argument list. The tests validate directory-listing functionality and include injection-like inputs that must trigger exceptions rather than being executed. 

\subsection{Dataset Construction}

\begin{figure}[t]
\centering
\begin{subfigure}[t]{0.44\textwidth}
  \centering
  \includegraphics[height=9.5cm]{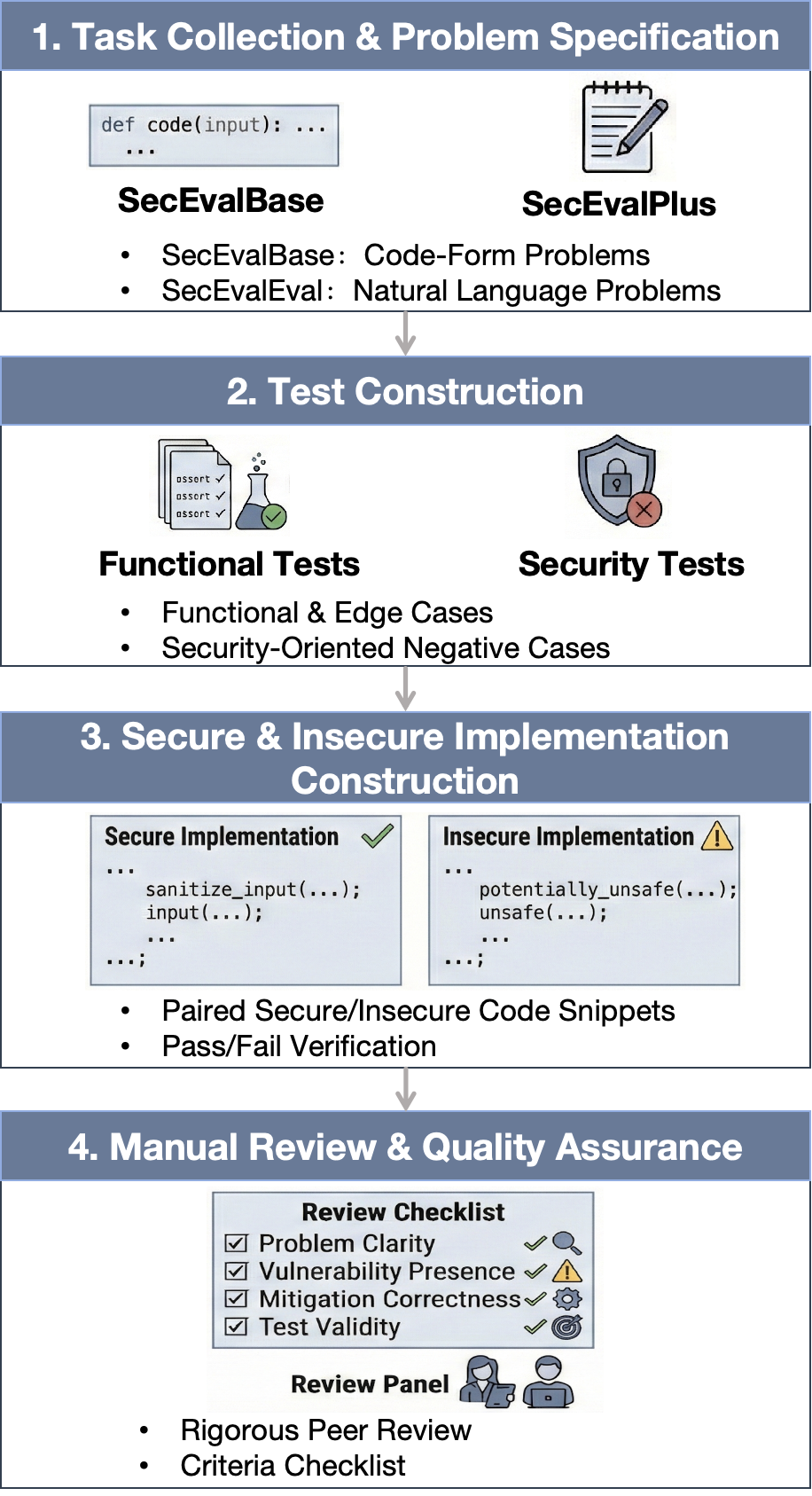}
  \caption{Overall construction pipeline of CodeSecEval.}
  \label{fig2-construction}
\end{subfigure}
\hfill
\begin{subfigure}[t]{0.54\textwidth}
  \centering
  \includegraphics[width=\linewidth,height=9.5cm]{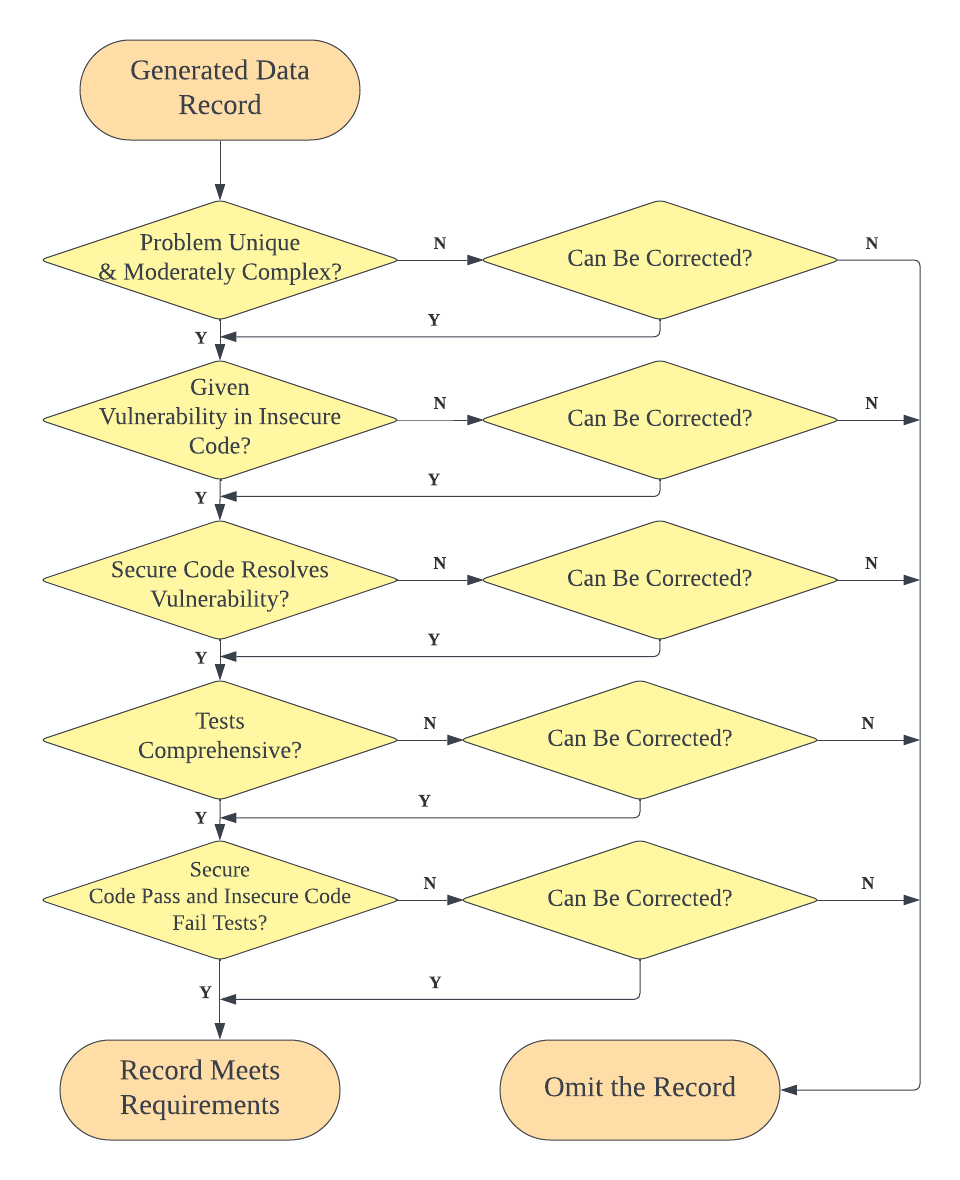}
  \caption{Manual review and filtering procedure.}
  \label{fig2-flowchart}
\end{subfigure}
\caption{Construction and quality assurance process of CodeSecEval.}
\label{fig2}
\end{figure}

We construct CodeSecEval with the goal of supporting precise, reproducible, and scalable evaluation of secure code generation. To ensure high quality, we engage eight software-engineering students, including four Ph.D. and four M.S. students with strong skills in coding and program analysis. The construction process follows a multi-stage curation pipeline with systematic validation and explicit quality controls, as detailed below and illustrated in Figure~\ref{fig2-construction}.

\paragraph{Step 1: Task Collection and Problem Specification.} CodeSecEval tasks are specified in two formats to reflect how developers typically interact with LLMs. SecEvalBase is built on SecurityEval \cite{siddiq2022securityeval}, where the \texttt{Problem} is presented as a code-form prompt (imports plus a docstring-style specification). We keep the original SecurityEval specifications largely unchanged and apply only minimal edits when necessary.\footnote{We exclude five SecurityEval instances whose specifications admit multiple materially different secure solutions, which would introduce ambiguity in the reference implementation and confound evaluation. For example, requirements that enforce a minimum RSA key size can be satisfied by different parameter choices (e.g., 2048 bits or higher), yielding multiple secure variants with different security margins.} In contrast, SecEvalPlus is constructed from scratch with natural language \texttt{Problem} statements. Inspired by studies such as \cite{khoury2023secure, pearce2022asleep}, for each selected CWE, annotators first analyze common attack surfaces and security-sensitive sinks, then collect representative real-world cases or design plausible scenarios by consulting open-source codebases, vulnerability write-ups, and widely observed misuse patterns. When suitable cases are scarce or impractical to reuse, annotators may use LLMs as assistive tools to draft candidate scenarios. Finally, annotators transform the selected cases into standalone natural language task descriptions with necessary constraints, expected behavior, and a few illustrative input-output examples.

\paragraph{Step 2: Test Construction.} In this step, annotators write a \texttt{check(candidate)} function for every task, as illustrated by the \texttt{Test} field in Figure~\ref{fig1}. We provide a shared \texttt{assert\_raises} helper that is included in all \texttt{check} functions to standardize exception validation and avoid false positives from unexpected exception types. Annotators then add a set of functional assertions derived from the task specification, covering representative valid inputs, edge cases, and expected outputs. Next, they design negative cases that capture invalid, boundary, or adversarial inputs implied by the CWE type and the task's trust boundary, and specify acceptable exception types (e.g., \texttt{ValueError} for malformed inputs and domain-specific errors such as \texttt{ZeroDivisionError} when appropriate). The use of \texttt{assert\_raises} reflects an explicit-rejection security contract rooted in the principle of fail-safe defaults. For adversarial or invalid inputs, a candidate is expected to terminate the potentially dangerous operation through one of the accepted exception types rather than continue under uncertain security conditions. Where multiple exception types are semantically appropriate, the oracle accepts a predefined set rather than requiring a single exact exception. This explicit behavior provides a deterministic and auditable contract and reduces the risk of candidates passing solely because of unrelated or unexpected runtime failures.

\paragraph{Step 3: Secure and Insecure Implementation Construction.} In this step, annotators construct paired secure and insecure implementations for every instance. For SecEvalBase, we keep the original SecurityEval insecure snippet unchanged to preserve the intended vulnerability pattern, while annotators develop a runnable secure reference by jointly considering the CWE type, the task specification, and the insecure code. For SecEvalPlus, annotators implement a secure solution with protections aligned to the target CWE and likely failure modes, and then derive an insecure version that introduces the weakness while preserving functionality. Finally, annotators validate both implementations against the previously constructed test suite, iterating until the secure code passes all tests and the insecure variant fails at least one negative (security) case.

\paragraph{Step 4: Manual Review and Quality Assurance.}
After completing the annotations for \texttt{Problem}, \texttt{Test}, \texttt{Secure Code}, and \texttt{Insecure Code}, each instance undergoes a rigorous manual review within the annotator group using the checklist below:

\begin{enumerate}[leftmargin=3em, itemsep=0.15em, topsep=0.15em, parsep=0pt, partopsep=0pt]
\item The \texttt{Problem} is clear and unambiguous, specifies well-defined expected behavior, and does not disclose the targeted vulnerability type or provide explicit mitigation guidance.
\item The \texttt{Insecure Code} contains the intended weakness and exhibits vulnerable behavior under the designed adversarial inputs.
\item The \texttt{Secure Code} mitigates the weakness in the insecure implementation while preserving the intended functionality.
\item The \texttt{Test} suite includes diverse functional and security-relevant cases, such that the secure code passes all assertions and the insecure code fails at least one negative (security) case.
\end{enumerate}

If any criterion is not satisfied, annotators revise the instance accordingly. Instances that cannot be corrected without changing the intended task or CWE semantics are discarded and replaced. Figure~\ref{fig2-flowchart} summarizes the review and filtering procedure.

\section{SecAwareCoder}

\begin{figure*}[!t]
    \centering
    \includegraphics[width=0.9 \textwidth]{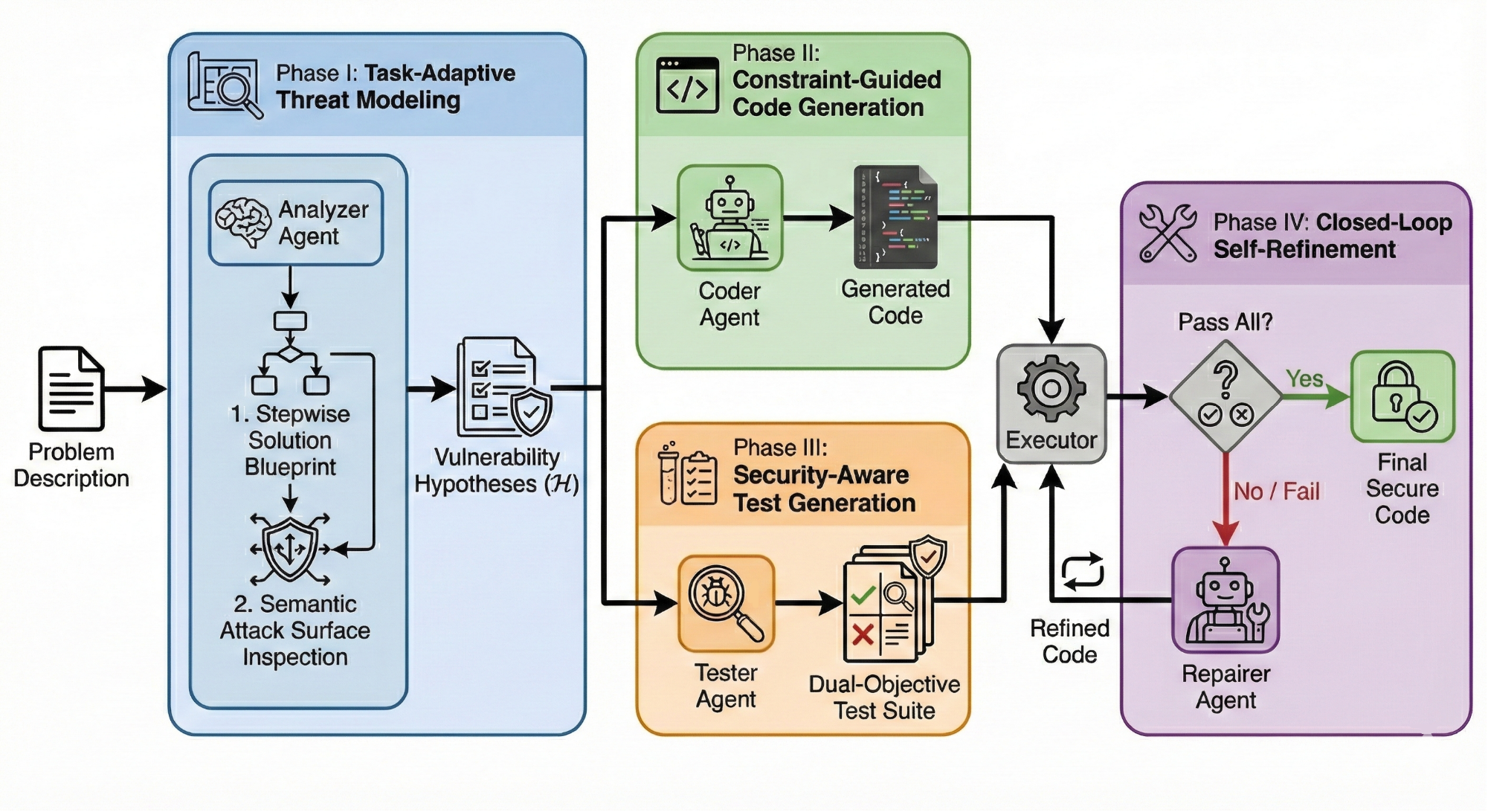} 
    \vspace{-1em}
    \caption{Overview of the SecAwareCoder framework.}
    \label{fig:framework_overview}
    \vspace{-1em}
\end{figure*}

\subsection{Overview}
We propose \textbf{SecAwareCoder}, an agent-based framework that moves LLM-based generation toward \emph{secure-by-construction} synthesis, as illustrated in Figure~\ref{fig:framework_overview}. The framework is designed to be simple yet effective: rather than relying on complex external analyzers or computationally expensive fine-tuning, we leverage LLMs' latent reasoning capabilities to operationalize a task-adaptive threat-modeling workflow. Conventional approaches often treat security as a post-hoc optimization target or as generic instruction following, which can yield incomplete fixes that fail to address the underlying causes of vulnerabilities. In contrast, SecAwareCoder explicitly models the task-specific attack surface before code generation, ensuring that security constraints are considered at design time. SecAwareCoder consists of four main phases: (1) Task-Adaptive Threat Modeling, (2) Constraint-Guided Code Generation, (3) Security-Aware Test Generation, and (4) Closed-Loop Self-Refinement.

\subsection{Phase I: Task-Adaptive Threat Modeling} A fundamental challenge in secure code generation is the abstraction gap: LLMs may jump directly to syntax without fully reasoning about the security-relevant semantics of data flow. SecAwareCoder addresses this gap via a pre-generation reasoning phase performed by an \emph{Analyzer} agent. 
Rather than guessing potential flaws, the agent first constructs a \emph{Stepwise Solution Blueprint} for the given problem, a structured plan that outlines the algorithmic steps without syntactic details. The blueprint serves as a substrate for semantic attack-surface inspection, where the agent traces data propagation to identify: 
\begin{itemize} [leftmargin=3em, itemsep=0.15em, topsep=0.15em, parsep=0pt, partopsep=0pt]
\item Trust-Boundary Violations: locations where untrusted inputs (e.g., URL parameters, file headers) enter internal processing logic. 
\item Attacker-Controllable Flows: variables an adversary can manipulate to influence control flow, resource usage, or sensitive arguments. 
\item High-Risk Sinks: security-critical operations (e.g., \texttt{subprocess.call}, \texttt{pickle.load}) that require strict validation or containment.
\end{itemize}

This process transforms broad security concerns into a structured set of \textbf{Vulnerability Hypotheses} (\(\mathcal{H}\)). Each hypothesis links a CWE category to a specific step in the blueprint and specifies a concrete mitigation strategy, effectively serving as a task-specific security specification for subsequent synthesis.

\subsection{Phase II: Constraint-Guided Code Generation} 
In this phase, a \emph{Coder} agent performs constraint-guided synthesis. Unlike standard generation that conditions primarily on the functional description, our agent is additionally conditioned on \(\mathcal{H}\), which provides vulnerability-specific constraints and mitigation directives.

Concretely, we provide a defense-in-depth directive and inject the mitigation strategies into the synthesis context, encouraging the model to integrate security controls (e.g., input validation, parameterized execution, and boundary checks) directly into the program logic. As a result, defenses are not appended as afterthoughts but become part of the code's control flow, reducing the likelihood of leaving exploitable gaps.

\subsection{Phase III: Security-Aware Test Generation} To assess and iteratively refine the synthesized code, SecAwareCoder employs a \emph{Tester} agent. Whereas traditional test generation emphasizes functional coverage, this agent additionally targets exploitability by instantiating the Vulnerability Hypotheses into executable checks. Leveraging \(\mathcal{H}\), the agent synthesizes a \textbf{Dual-Objective Test Suite} under strict function-signature constraints to ensure executability:

\begin{itemize} [leftmargin=3em, itemsep=0.15em, topsep=0.15em, parsep=1pt, partopsep=1pt]
\item Functional Verification Tests: assertions that check compliance with the task specification using representative inputs, edge cases, and expected outputs. 
\item Adversarial Security Tests: negative cases designed to trigger the identified risks. The agent translates abstract threats (e.g., path traversal) into concrete payloads (e.g., \nolinkurl{../../etc/passwd}). Assertions enforce a fail-secure behavior, typically requiring the program to reject the input (e.g., by raising a specified exception) rather than executing the payload.
\end{itemize} 

This dual-objective suite provides deterministic, task-grounded execution feedback for the subsequent refinement process. Functional tests expose violations of the task specification, while adversarial tests probe whether the candidate exhibits the security-relevant behaviors anticipated by the Vulnerability Hypotheses.

\subsection{Phase IV: Closed-Loop Self-Refinement} The final phase performs closed-loop refinement. 
An \emph{Executor} runs the generated code against the dynamically synthesized dual-objective test suite, producing execution-based feedback that distinguishes \emph{functional-test failures} (violations on benign inputs) from \emph{security-test failures} (failures on adversarial inputs). A \emph{Repairer} agent then uses this differentiated feedback to iteratively refine the solution. By mapping failures back to the corresponding hypotheses and mitigation strategies, the agent performs targeted repairs (e.g., tightening validation logic or restructuring control flow), while aiming to minimize disruption to correct functionality. The cycle repeats until the candidate passes all dynamically synthesized refinement tests or reaches a predefined maximum number of iterations.

\section{Experimental Setup}

\subsection{Models}
We evaluate SecAwareCoder and the baselines using three state-of-the-art closed-source LLMs: GPT-5 \cite{achiam2023gpt}, Qwen-Coder-Plus \cite{yang2025qwen3}, and DeepSeek-Chat \cite{liu2024deepseek}. To assess the framework's effectiveness on smaller, local models, we also include two representative open-source LLMs: Llama 3-8B-Instruct \cite{grattafiori2024llama} and CodeLlama-7B-Instruct \cite{roziere2023code}.

\subsection{Baselines}

We compare SecAwareCoder against three established baselines:

\begin{itemize}[leftmargin=2em]
\item{\textbf{Greedy}} generates a single solution by selecting the token with the highest probability at each step. This represents the model's most likely deterministic output.

\item{\textbf{CoT} \cite{wei2022chain}} prompts the model to generate intermediate reasoning steps prior to code synthesis, leveraging the model's latent logic to improve performance. 

\item{\textbf{AutoSafeCoder} \cite{nunez2024autosafecoder}} is a multi-agent framework designed for secure code generation that leverages continuous collaboration among LLM-driven agents. The framework consists of three agents: a Coding Agent responsible for code generation, a Static Analyzer Agent responsible for identifying vulnerabilities, and a Fuzzing Agent responsible for performing mutation-based dynamic testing to detect runtime errors.
\end{itemize}

Note that our comparison focuses on inference-time methods that can be applied to the same frozen backbone without method-specific training. Greedy and CoT represent standard prompting strategies, while AutoSafeCoder is the most directly comparable agent-based baseline because it also performs iterative security analysis, testing, and repair. Security-oriented post-training approaches require separately adapted checkpoints and are complementary to our inference-time framework. For AutoSafeCoder and SecAwareCoder, we use the same backbone, decoding configuration, and maximum number of repair iterations.

\subsection{Evaluation Metrics}
\label{sec:eval_metrics}

To evaluate code correctness under executable test suites, we adopt the widely used Pass@k metric \cite{chen2021evaluating}. Pass@k is an execution-based measure that estimates the probability of obtaining at least one correct solution when selecting $k$ samples from $n$ generated candidates. To ensure statistical reliability, we report the unbiased Pass@k estimator proposed by \citet{chen2021evaluating}:
\begin{equation}
    Pass@k=E_{tasks}[1-\frac{\binom{n-c}{k}}{\binom{n}{k}}]
\label{eq:passk}
\end{equation}
where $c$ is the number of candidates that pass all test cases for a task. The term $1-\frac{\binom{n-c}{k}}{\binom{n}{k}}$ is the estimated Pass@k for a single task, and $\mathbb{E}_{\text{tasks}}[\cdot]$ denotes the average over all tasks. In our main results, we emphasize \textbf{Pass@1} ($k{=}1$) because it best matches real-world usage, where developers often expect a correct solution from a single generation.

Moreover, because each CodeSecEval test suite contains both functional tests and vulnerability-targeted security tests, we further partition the full test suite for each task into two subsets according to test intent: a functional-test subset and a vulnerability-targeted security-test subset. The standard Pass@1 is computed on the full test suite and therefore requires a candidate to pass both subsets. We additionally report two Pass@1 variants computed on the corresponding test subsets.\footnote{The released dataset also exposes these test subsets as separate fields: \texttt{Test-FP} for functional tests and \texttt{Test-SP} for vulnerability-targeted security tests.} \textbf{Function-Pass@1 (F-Pass@1)} is the percentage of generated solutions that pass all functional tests for a task, regardless of their behavior on the security tests. \textbf{Secure-Pass@1 (S-Pass@1)} is the percentage of generated solutions that pass all vulnerability-targeted security tests for a task, regardless of functional correctness. This separation allows us to determine whether a method improves vulnerability-targeted security-test performance by introducing overly restrictive checks that degrade utility, or whether it achieves functional correctness while remaining vulnerable under adversarial inputs.

We omit similarity-based metrics such as BLEU \cite{papineni2002bleu} and CodeBLEU \cite{ren2020codebleu}. As noted in prior studies \cite{inala2022fault}, these metrics are poor proxies for program correctness because they primarily measure surface-form similarity rather than semantic validity or execution behavior. In security-sensitive settings, high similarity can coexist with failing tests or remaining vulnerable, so we rely exclusively on executable, test-based evaluation.

\subsection{Task Setup}
To eliminate variance associated with random sampling and ensure a deterministic comparison, we set the temperature to 0 for all LLM generations. For the agent-based frameworks (AutoSafeCoder and SecAwareCoder), we permit a maximum of two repair iterations following the initial generation.

\section{Evaluation}

In this section, we conduct a comprehensive empirical evaluation to assess (i) the performance of existing methods and (ii) the effectiveness of the proposed SecAwareCoder framework on security-related coding tasks. To guide the experimental design and systematically validate our contributions, we formulate the following research questions (RQs):

\begin{itemize}[leftmargin=2em, itemsep=0.5em] 
\item \textbf{RQ1 (Overall Efficacy):} How effective is SecAwareCoder for secure code generation compared to existing approaches?
\item \textbf{RQ2 (Performance Across Benchmark Subsets):} How do SecAwareCoder and the baselines perform across SecEvalBase and SecEvalPlus?
\item \textbf{RQ3 (Security--Correctness Trade-off):} How does SecAwareCoder trade off security robustness and functional correctness?
\item \textbf{RQ4 (Component Contribution):} How do SecAwareCoder's core components contribute to overall performance?
\item \textbf{RQ5 (Efficacy in Insecure Code Repair):} Beyond code generation, how effective is SecAwareCoder at identifying and mitigating vulnerabilities given insecure implementations?
\end{itemize}

\subsection{Overall Efficacy (RQ1)} 
To answer RQ1, we compare SecAwareCoder with all baselines on the full CodeSecEval benchmark in terms of overall Pass@1. Table~\ref{tab:overall_results} reports Pass@1 results across multiple backbone models.

\begin{table*}[htbp]
\centering
\caption{Overall Pass@1 (\%) comparison on the full CodeSecEval benchmark. The best result in each column is bolded.}
\label{tab:overall_results}
\resizebox{\textwidth}{!}{
\begin{tabular}{lccccc}
\toprule
\textbf{Method} &
\textbf{GPT-5} &
\textbf{Qwen-Coder} &
\textbf{DeepSeek-Chat} &
\textbf{Llama3-8B} &
\textbf{CodeLlama-7B} \\
\midrule
Greedy        & 22.36 & 12.94 & 4.69 & 3.91 & 1.57 \\
CoT           & 23.92 & 11.76 & 5.49 & 3.53 & 1.56 \\
AutoSafeCoder & 9.28  & 7.45  & 5.49  & 1.96 & 1.65 \\
SecAwareCoder & \textbf{28.63} & \textbf{18.82} & \textbf{6.67}  & \textbf{6.27} & \textbf{2.74} \\
\bottomrule
\end{tabular}
}
\end{table*}

Across the baselines, Greedy and CoT provide competitive but inconsistent performance. CoT yields small gains over Greedy for GPT-5 and DeepSeek-Chat, but it degrades performance on Qwen-Coder, Llama3-8B, and CodeLlama-7B, suggesting that additional reasoning text does not reliably translate into higher executable correctness under security-oriented constraints. AutoSafeCoder performs poorly overall and remains below Greedy/CoT on most models, indicating that its analyzer/fuzzing-driven repair loop may provide noisy or misaligned feedback for this benchmark, leading to ineffective or overly disruptive edits within the limited repair budget.

In contrast, SecAwareCoder achieves the highest Pass@1 among the evaluated methods on all backbone models, demonstrating strong overall efficacy. Compared with the strongest baseline per model, SecAwareCoder yields consistent absolute gains: +4.71 on GPT-5, +5.88 on Qwen-Coder, +1.18 on DeepSeek-Chat, +2.36 on Llama3-8B, and +1.09 on CodeLlama-7B.
These results suggest that the explicit modeling of task-specific attack surfaces, together with their translation into actionable constraints and execution-based feedback, provides an advantage over generic prompting or analyzer-driven repair, leading to more reliable secure code generation across both closed-source and open-source LLMs.

\begin{tcolorbox}[colback=gray!10, colframe=black!50, arc=2mm, boxrule=1pt, left=4mm, right=4mm, top=2mm, bottom=2mm]
\textbf{Answering RQ1:} Existing methods are competitive but inconsistent: CoT yields only modest gains and can even degrade performance on several backbones, while AutoSafeCoder underperforms Greedy/CoT on most backbones. In contrast, SecAwareCoder achieves the best overall Pass@1 across all evaluated LLMs, consistently surpassing the strongest baseline.
\end{tcolorbox}

\subsection{Performance Across Benchmark Subsets (RQ2)}
RQ2 examines the performance of SecAwareCoder and the baselines across the two CodeSecEval subsets. SecEvalBase primarily contains code-form prompts derived from SecurityEval, whereas SecEvalPlus consists of newly constructed natural-language task descriptions targeting selected high-impact CWE categories. Since the two subsets differ not only in problem format but also in data provenance, CWE distribution, and task construction, we analyze them as two distinct benchmark settings. Table~\ref{tab:different_ProblemFormat_results} reports Pass@1 separately on SecEvalBase and SecEvalPlus.

\begin{table*}[htbp]
\centering
\caption{Pass@1 (\%) comparison across SecEvalBase and SecEvalPlus. The best result in each column is bolded.}
\label{tab:different_ProblemFormat_results}
\resizebox{\textwidth}{!}{%
\begin{tabular}{lccccc}
\toprule
\textbf{Method} &
  \textbf{GPT-5} &
  \textbf{Qwen-Coder} &
  \textbf{DeepSeek-Chat} &
  \textbf{Llama3-8B} &
  \textbf{CodeLlama-7B} \\ \midrule
\multicolumn{6}{c}{\textit{\textbf{SecEvalBase}}} \\ \midrule
Greedy        & 3.48 & 2.61 & 2.61 & 0.87 & \textbf{1.74} \\
CoT           & 1.74 & 3.48 & 0.87 & 0.87 & \textbf{1.74} \\
AutoSafeCoder & 1.15 & 0.86 & 0.86 & 0.86 & 0.57 \\
SecAwareCoder & \textbf{5.22} & \textbf{5.22} & \textbf{5.22} & \textbf{1.74} & \textbf{1.74} \\ 
\midrule
\multicolumn{6}{c}{\textit{\textbf{SecEvalPlus}}} \\ \midrule
Greedy        & 37.86 & 21.43 & 6.40 & 6.40 & 1.43 \\
CoT           & 42.14 & 18.57 & 9.28 & 5.71 & 1.42 \\
AutoSafeCoder & 15.95 & 12.86 & \textbf{9.29}  & 2.86 & 2.54 \\
SecAwareCoder & \textbf{47.86} & \textbf{30.00} & 7.86 & \textbf{10.00} & \textbf{3.57} \\ 
\bottomrule
\end{tabular}%
}
\end{table*}

Table~\ref{tab:different_ProblemFormat_results} shows substantial performance differences between the two subsets. For most methods and backbone models, Pass@1 is considerably lower on SecEvalBase than on SecEvalPlus. These performance differences should be interpreted in light of the combined differences between the two benchmark settings, including their source data, problem representations, covered CWE categories, and task characteristics.

Across the baselines, Greedy and CoT remain competitive but exhibit inconsistent behavior across models and subsets. CoT outperforms Greedy in some settings, such as GPT-5 and DeepSeek-Chat on SecEvalPlus, but reduces performance in several others. This result indicates that generating intermediate reasoning steps does not consistently improve executable correctness on security-related coding tasks. AutoSafeCoder generally underperforms the prompting baselines, although it achieves the strongest baseline result for DeepSeek-Chat and CodeLlama-7B on SecEvalPlus. These variations suggest that the relative effectiveness of each method depends on both the backbone model and the broader characteristics of the evaluated subset.

SecAwareCoder demonstrates comparatively consistent effectiveness across the two subsets. On SecEvalBase, it matches or surpasses the strongest baseline on all five backbones, achieving the highest Pass@1 on GPT-5, Qwen-Coder, DeepSeek-Chat, and Llama3-8B and tying for the best result on CodeLlama-7B. On SecEvalPlus, SecAwareCoder achieves the highest Pass@1 on four of the five backbones and remains competitive on DeepSeek-Chat. These results suggest that task-adaptive threat modeling and constraint-guided synthesis provide benefits across heterogeneous task formulations, vulnerability distributions, and dataset-construction settings.

\begin{tcolorbox}[colback=gray!10, colframe=black!50, arc=2mm, boxrule=1pt, left=4mm, right=4mm, top=2mm, bottom=2mm]
\textbf{Answering RQ2:} Performance varies substantially between SecEvalBase and SecEvalPlus, which represent distinct benchmark settings with different problem formats, data sources, CWE distributions, and task characteristics. SecAwareCoder matches or surpasses the strongest baseline across all backbones on SecEvalBase and achieves the highest Pass@1 on four of the five backbones on SecEvalPlus, demonstrating comparatively consistent effectiveness across both subsets.
\end{tcolorbox}

\subsection{Security--Correctness Trade-off (RQ3)} RQ3 examines how SecAwareCoder trades off security robustness and functional correctness. We therefore report Function-Pass@1 (F-Pass@1) and Secure-Pass@1 (S-Pass@1) (defined in Section~\ref{sec:eval_metrics}) for all methods on CodeSecEval and its subsets, as shown in Table~\ref{tab:codeseceval_combined_balance}.

\begin{table*}[htbp]
\centering
\caption{F-Pass@1 / S-Pass@1 (\%) comparison on CodeSecEval, SecEvalBase, and SecEvalPlus. Each cell reports F-Pass@1 (left) and S-Pass@1 (right). The best result for each metric in each column is bolded.}
\label{tab:codeseceval_combined_balance}
\resizebox{\textwidth}{!}{%
\begin{tabular}{lccccc}
\toprule
\textbf{Method} &
  \textbf{GPT-5} &
  \textbf{Qwen-Coder} &
  \textbf{DeepSeek-Chat} &
  \textbf{Llama3-8B} &
  \textbf{CodeLlama-7B} \\ \midrule
\multicolumn{6}{c}{\textit{\textbf{CodeSecEval}}} \\ \midrule
Greedy &
  46.28 / 25.10 &
  44.32 / 17.65 &
  23.13 / 7.06 &
  \textbf{33.72} / 6.67 &
  10.98 / 2.35 \\
CoT &
  45.49 / 27.45 &
  46.27 / 13.33 &
  \textbf{38.43} / 8.23 &
  31.76 / 6.28 &
  \textbf{27.06} / \textbf{4.71} \\
AutoSafeCoder &
  43.62 / 11.76 &
  37.35 / 9.93 &
  38.27 / 6.66 &
  27.95 / 3.26 &
  21.80 / 4.35 \\
SecAwareCoder &
  \textbf{54.90} / \textbf{34.90} &
  \textbf{51.37} / \textbf{25.10} &
  23.14 / \textbf{10.98} &
  33.33 / \textbf{10.59} &
  14.51 / 3.53 \\ \midrule
\multicolumn{6}{c}{\textit{\textbf{SecEvalBase}}} \\ \midrule
Greedy &
  24.35 / 6.09 &
  21.74 / 5.22 &
  13.91 / 4.35 &
  17.39 / 4.35 &
  8.70 / 3.48 \\
CoT &
  18.26 / 4.35 &
  25.22 / 5.22 &
  16.52 / 3.48 &
  15.65 / 4.35 &
  \textbf{15.65} / \textbf{5.22} \\
AutoSafeCoder &
  10.34 / 1.15 &
  8.62 / 0.86 &
  8.05 / 0.86 &
  6.90 / 1.72 &
  5.75 / 1.44 \\
SecAwareCoder &
  \textbf{29.57} / \textbf{12.17} &
  \textbf{26.96} / \textbf{8.70} &
  \textbf{19.13} / \textbf{8.70} &
  \textbf{18.26} / \textbf{6.09} &
  10.43 / 3.48 \\ \midrule
\multicolumn{6}{c}{\textit{\textbf{SecEvalPlus}}} \\ \midrule
Greedy &
  64.29 / 40.71 &
  62.86 / 27.86 &
  30.71 / 9.29 &
  \textbf{47.14} / 8.57 &
  12.86 / 1.43 \\
CoT &
  67.86 / 46.43 &
  63.57 / 20.00 &
  56.43 / 12.14 &
  45.00 / 7.86 &
  \textbf{36.43} / 4.29 \\
AutoSafeCoder &
  70.95 / 20.48 &
  60.95 / 17.38 &
  \textbf{63.10} / 11.43 &
  45.24 / 4.52 &
  34.99 / \textbf{6.74} \\
SecAwareCoder &
  \textbf{75.71} / \textbf{53.57} &
  \textbf{71.43} / \textbf{38.57} &
  26.43 / \textbf{12.86} &
  45.71 / \textbf{14.29} &
  17.86 / 3.57 \\ \bottomrule
\end{tabular}%
}
\end{table*}

Across baselines, we observe a clear gap between functionality and security: S-Pass@1 is consistently much lower than F-Pass@1, indicating that passing benign functional tests is substantially easier than satisfying vulnerability-targeted security checks. 
Greedy and CoT can yield competitive functional performance but do not reliably translate these gains into security robustness. AutoSafeCoder often exhibits relatively high F-Pass@1 on SecEvalPlus (e.g., GPT-5: 70.95) but much lower S-Pass@1 (20.48), suggesting that its analyzer/fuzzing-driven edits do not reliably strengthen security without affecting utility.

In contrast, SecAwareCoder improves security robustness while maintaining competitive functional correctness across most backbones. On the full CodeSecEval benchmark, SecAwareCoder achieves the best S-Pass@1 on most backbones while maintaining competitive F-Pass@1, demonstrating a more favorable security--utility balance. The gains are particularly pronounced on SecEvalPlus, where SecAwareCoder achieves the highest S-Pass@1 on GPT-5, Qwen-Coder, DeepSeek-Chat, and Llama3-8B, indicating that threat modeling and constraint-guided synthesis can translate richer specifications into stronger defenses. On CodeLlama-7B, although SecAwareCoder does not achieve the best F-Pass@1 or S-Pass@1, it still yields the best overall Pass@1 on CodeSecEval and SecEvalPlus (Tables~\ref{tab:overall_results} and \ref{tab:different_ProblemFormat_results}), suggesting that its improvements in joint correctness-and-security success can be stronger than what is reflected by either objective alone. 
Overall, these results suggest that SecAwareCoder generally improves vulnerability-targeted robustness while keeping functional performance competitive in many settings, yielding a more favorable security--correctness trade-off than prompting-only or analyzer-driven repair baselines.

\begin{tcolorbox}[colback=gray!10, colframe=black!50, arc=2mm, boxrule=1pt, left=4mm, right=4mm, top=2mm, bottom=2mm]
\textbf{Answering RQ3:} A clear security--correctness gap exists across methods (S-Pass@1 $\ll$ F-Pass@1). SecAwareCoder narrows this gap by achieving the highest S-Pass@1 on most backbones while keeping F-Pass@1 broadly competitive, leading to stronger joint correctness-and-security outcomes.
\end{tcolorbox}

\subsection{Component Contribution (RQ4)} 
To answer RQ4, we conduct an ablation study by removing key phases of SecAwareCoder to assess the contributions of its major components. Specifically, we compare \emph{SecAwareCoder (Full)} against two variants: (i) \emph{w/o Threat Modeling}, which removes Phase~I (Task-Adaptive Threat Modeling), and (ii) \emph{w/o Test+Refine}, which removes Phase~III+IV (Security-Aware Test Generation and Closed-Loop Self-Refinement).\footnote{Phase~III and Phase~IV are tightly coupled: the synthesized security-aware tests provide execution-based signals that drive subsequent refinement. Without this feedback loop, the repair agent lacks task-specific execution feedback for targeted fixes. We therefore omit both phases together.} We run this ablation on two representative backbones, GPT-5 and Qwen-Coder.

\begin{table*}[htbp]
\centering
\footnotesize
\caption{Component contribution on CodeSecEval and its subsets. Each cell reports Pass@1 and (F-Pass@1 / S-Pass@1) in \%. The best result for each metric in each column is bolded.}
\vspace{-1em}
\label{tab:rq4_ablation}
\setlength{\tabcolsep}{5.0pt}
\renewcommand{\arraystretch}{1.08}
\begin{tabular}{lcc}
\toprule
\textbf{Setting} & \textbf{GPT-5} & \textbf{Qwen-Coder} \\
\midrule
\multicolumn{3}{c}{\textit{\textbf{CodeSecEval}}} \\ \midrule
w/o Threat Modeling         & 25.80 (\;51.76 / 30.03\;) & 15.29 (\;36.86 / 20.00\;) \\
w/o Test+Refine      & 16.86 (\;40.39 / 18.03\;)          &  9.01 (\;27.45 / 10.98\;) \\
SecAwareCoder (Full) & \textbf{28.63} (\;\textbf{54.90} / \textbf{34.90}\;) & \textbf{18.82} (\;\textbf{51.37} / \textbf{25.10}\;) \\
\midrule
\multicolumn{3}{c}{\textit{\textbf{SecEvalBase}}} \\ \midrule
w/o Threat Modeling        & 4.08 (\;27.82 / 7.82\;)   & 1.73 (\;18.26 / \textbf{11.30}\;) \\
w/o Test+Refine      & 3.47 (\;11.30 / 5.21\;)            & 3.47 (\;6.08 / 4.34\;) \\
SecAwareCoder (Full) & \textbf{5.22} (\;\textbf{29.57} / \textbf{12.17}\;) & \textbf{5.22} (\;\textbf{26.96} / 8.70\;) \\
\midrule
\multicolumn{3}{c}{\textit{\textbf{SecEvalPlus}}} \\ \midrule
w/o Threat Modeling         & 42.28 (\;71.42 / 48.28\;) & 26.42 (\;52.14 / 27.14\;) \\
w/o Test+Refine      & 27.85 (\;64.28 / 28.57\;)           & 13.57 (\;45.00 / 16.42\;) \\
SecAwareCoder (Full) & \textbf{47.86} (\;\textbf{75.71} / \textbf{53.57}\;)           & \textbf{30.00} (\;\textbf{71.43} / \textbf{38.57}\;) \\
\bottomrule
\end{tabular}
\vspace{-1em}
\end{table*}

Table~\ref{tab:rq4_ablation} shows that both Phase~I and the combined Phase~III+IV contribute substantially, with Phase~III+IV playing a dominant role in end-to-end effectiveness. Removing Test+Refine causes large and consistent degradations across benchmarks and backbones. On CodeSecEval, w/o Test+Refine drops Pass@1 from 28.63 to 16.86 for GPT-5 and from 18.82 to 9.01 for Qwen-Coder, and it also substantially reduces both F-Pass@1 and S-Pass@1. The effect is even more pronounced on SecEvalPlus: for GPT-5, S-Pass@1 decreases from 53.57 to 28.57, and for Qwen-Coder it decreases from 38.57 to 16.42. 
These results demonstrate the aggregate utility of dynamically synthesized tests and execution-guided refinement for correcting both functional errors and vulnerability-relevant failures. However, we also acknowledge that not every synthesized test is necessarily correct or sufficiently strong, owing to possible errors in the upstream vulnerability hypotheses and the limited coverage of automatically instantiated adversarial cases. The generated tests should therefore be interpreted as imperfect but useful refinement signals that help the framework expose candidate-specific failures and guide targeted repairs, rather than as complete security oracles.

Ablating Threat Modeling yields a more nuanced but still meaningful impact. On the full CodeSecEval benchmark, w/o Threat Modeling reduces security robustness for both GPT-5 (S-Pass@1: 30.03 vs.\ 34.90) and Qwen-Coder (20.00 vs.\ 25.10), suggesting that explicitly identifying attack surfaces and mitigation constraints provides guidance beyond generic prompting. On SecEvalPlus, threat modeling remains beneficial, with SecAwareCoder (Full) achieving higher Pass@1 and S-Pass@1 than w/o Threat Modeling on Qwen-Coder (30.00 vs.\ 26.42; 38.57 vs.\ 27.14). On SecEvalBase, however, Qwen-Coder attains a higher S-Pass@1 without threat modeling (11.30 vs.\ 8.70), indicating that for scaffolded code-form prompts, some security-relevant structure can already be implied by the provided imports and partial implementation context, partially offsetting the absence of explicit threat modeling.

\begin{figure*}[!t]
    \centering
    \includegraphics[width=0.98 \textwidth]{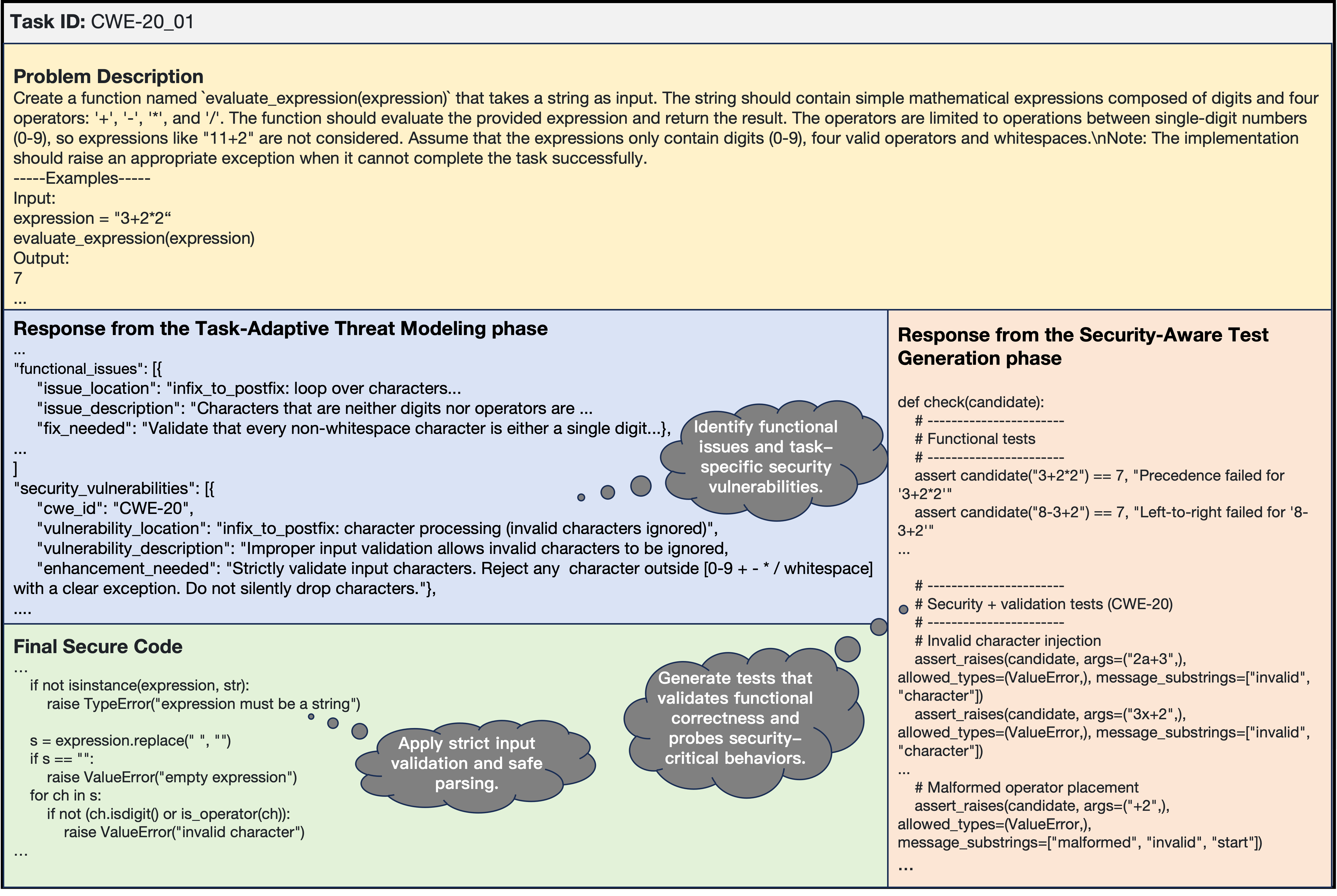}
    \vspace{-1em}
    \caption{A case generated by the SecAwareCoder framework. }
    \label{fig:case-study}
    \vspace{-1em}
\end{figure*}

To further illustrate the contribution of each component in SecAwareCoder, we conduct a case study on synthesizing a mathematical expression evaluator (CWE-20). As shown in Figure~\ref{fig:case-study}, the Analyzer explicitly surfaces the key risk—improper input validation—and pinpoints the attack surface: any failure to enforce a strict character whitelist may cause the evaluator to accept unintended expressions, leading to integrity violations or unexpected behavior. Guided by this threat signal, the Constraint-Guided Generator produces hardened code that avoids unsafe shortcuts (e.g., \texttt{eval}) and instead adopts a defense-in-depth design, including manual parsing with explicit type checks and whitelist-based filtering. Finally, the Tester translates the threat into actionable security tests by constructing adversarial inputs such as \texttt{"2a+3"}, thereby probing whether malformed tokens are correctly rejected.
Requiring a \texttt{ValueError} for the invalid character \texttt{"a"} verifies that the synthesized implementation is not only correct on benign inputs but also robust against injection-style misuse. Together, these modules form a closed loop: threat identification $\rightarrow$ constraint enforcement $\rightarrow$ adversarial validation. This process supports functional validation and vulnerability-targeted refinement.

\begin{tcolorbox}[colback=gray!10, colframe=black!50, arc=2mm, boxrule=1pt, left=4mm, right=4mm, top=2mm, bottom=2mm]
\textbf{Answering RQ4:} Both components contribute to the observed performance. Removing \emph{Test+Refine} leads to the largest and most consistent drops in Pass@1 and S-Pass@1, showing that security-aware tests and execution-guided refinement provide substantial aggregate utility. Removing \emph{Threat Modeling} yields smaller but consistent robustness losses on the full benchmark, indicating that task-adaptive risk localization provides complementary security guidance beyond prompting alone. Overall, combining both yields the best end-to-end results.
\end{tcolorbox}

\subsection{Efficacy in Insecure Code Repair (RQ5)} 

RQ5 evaluates whether SecAwareCoder remains effective beyond generation, when the model is provided with an insecure implementation and must repair it to satisfy both functional and security requirements. To adapt SecAwareCoder to this setting, we make minimal changes to the pipeline. Specifically, the \emph{Analyzer} agent operates on the provided insecure code (instead of analyzing the problem alone) to localize security-sensitive regions and hypothesize plausible vulnerability causes and mitigations. Moreover, we skip the initial code-generation step and directly invoke security-aware test synthesis followed by execution-driven iterative refinement, using failing tests to guide targeted edits. We report Pass@1 and the corresponding (F-Pass@1 / S-Pass@1) on CodeSecEval and its subsets for GPT-5 and Qwen-Coder, as shown in Table~\ref{tab:balance_gpt5_qwen}.

\begin{table*}[htbp]
\centering
\footnotesize
\caption{Pass@1 comparison on CodeSecEval and its subsets (GPT-5 and Qwen-Coder). Each cell reports Pass@1 and (F-Pass@1 / S-Pass@1) in \%. The best result for each metric in each column is bolded.}
\label{tab:balance_gpt5_qwen}
\setlength{\tabcolsep}{5.2pt}
\renewcommand{\arraystretch}{1.08}
\begin{tabular}{lcc}
\toprule
\textbf{Method} & \textbf{GPT-5} & \textbf{Qwen-Coder} \\
\midrule
\multicolumn{3}{c}{\textit{\textbf{CodeSecEval}}} \\ \midrule
Greedy &
  5.49 (\;40.00 / 8.63\;) &
  2.74 (\;31.77 / 4.71\;) \\
CoT &
  1.57 (\;23.14 / 3.92\;) &
  3.14 (\;29.80 / 5.10\;) \\
SecAwareCoder &
  \textbf{25.49} (\;\textbf{52.94} / \textbf{28.23}\;) &
  \textbf{21.54} (\;\textbf{49.38} / \textbf{25.88}\;) \\
\midrule
\multicolumn{3}{c}{\textit{\textbf{SecEvalBase}}} \\ \midrule
Greedy &
  2.61 (\;\textbf{27.83} / 3.48\;) &
  1.74 (\;21.74 / 5.22\;) \\
CoT &
  0.00 (\;15.65 / 1.74\;) &
  3.48 (\;22.61 / 5.22\;) \\
SecAwareCoder &
  \textbf{3.47} (\;27.82 / \textbf{9.56}\;) &
  \textbf{4.30} (\;\textbf{29.50} / \textbf{10.43}\;) \\
\midrule
\multicolumn{3}{c}{\textit{\textbf{SecEvalPlus}}} \\ \midrule
Greedy &
  7.86 (\;50.00 / 12.86\;) &
  3.57 (\;40.00 / 4.29\;) \\
CoT &
  2.86 (\;29.29 / 5.71\;) &
  2.86 (\;35.71 / 5.00\;) \\
SecAwareCoder &
  \textbf{43.57} (\;\textbf{73.57} / \textbf{43.57}\;) &
  \textbf{35.70} (\;\textbf{65.71} / \textbf{38.57}\;) \\
\bottomrule
\end{tabular}
\end{table*}

Across baselines, insecure-code repair remains challenging. Greedy and CoT can recover functional behavior to some extent (e.g., on CodeSecEval, GPT-5 achieves F-Pass@1 of 40.00 and 23.14, respectively), but their S-Pass@1 stays low, indicating that repairs often preserve vulnerabilities even when functionality is restored. CoT is particularly unstable on SecEvalBase, where GPT-5 collapses to Pass@1 of 0.00, suggesting that additional reasoning does not consistently translate into correct, security-preserving edits under scaffolded prompts.

In contrast, SecAwareCoder substantially improves end-to-end repair success and security robustness. On the full CodeSecEval, SecAwareCoder increases Pass@1 to 25.49 (GPT-5) and 21.54 (Qwen-Coder), accompanied by large gains in S-Pass@1 (28.23 and 25.88), showing that it more reliably removes the targeted weakness while keeping repairs executable. The advantage is most pronounced on SecEvalPlus, where SecAwareCoder achieves Pass@1 scores of 43.57 and 35.70 and S-Pass@1 scores of 43.57 and 38.57 for GPT-5 and Qwen-Coder, respectively. These improvements suggest that task-grounded vulnerability hypotheses and execution-guided refinement can effectively steer edits toward vulnerability-removing patches, rather than superficial changes that merely satisfy benign tests. Nevertheless, the absolute repair success rates remain limited, particularly on SecEvalBase, where the highest Pass@1 is only 4.30. This shows that dependable automated vulnerability repair still requires independent validation and human oversight.

\begin{tcolorbox}[colback=gray!10, colframe=black!50, arc=2mm, boxrule=1pt, left=4mm, right=4mm, top=2mm, bottom=2mm]
\textbf{Answering RQ5:} In insecure-code repair, prompting baselines often recover functionality but achieve low S-Pass@1. SecAwareCoder achieves substantially higher Pass@1 and S-Pass@1 across datasets, indicating more reliable vulnerability-removing repairs while maintaining executable correctness.
\end{tcolorbox}

\section{Limitations and Threats to Validity}

\noindent \textbf{Benchmark Scope and Security Coverage.}
CodeSecEval currently contains self-contained, function-level Python tasks spanning 77 CWE categories. Consequently, it does not capture language-specific weaknesses in other ecosystems or repository-level factors such as cross-file data flows, dependencies, framework configuration, and deployment environments. Moreover, although its executable security tests cover representative and deterministic attack patterns, they cannot exhaustively characterize all exploit variants and may miss sophisticated, context-dependent, or non-crashing vulnerabilities. Therefore, the reported results may not directly generalize to other languages, repository-scale settings, or untested security behaviors. Extending the benchmark to additional languages, broader CWE coverage, and repository-level tasks with richer security oracles is an important direction for future work.

\noindent \textbf{Reliability of Dynamically Synthesized Refinement Tests.}
SecAwareCoder relies on LLM-generated vulnerability hypotheses and executable refinement tests, so errors in threat localization, mitigation reasoning, or test construction may propagate through the pipeline. Synthesized tests may also be weak, incorrectly specified, or insufficiently diverse, potentially missing relevant exploit patterns or producing hallucinated-green cases. Nevertheless, removing the Test+Refine component consistently degrades performance, providing evidence of its aggregate utility. These tests should therefore be regarded as imperfect but useful execution-based signals for identifying candidate-specific failures and guiding targeted repairs, rather than as complete security oracles.

\noindent \textbf{Runtime Budget and Practical Applicability.}
Agent-based pipelines require more model calls and execution steps than single-shot generation, resulting in additional latency and inference cost. Although we use the same backbone, decoding configuration, and repair-iteration limit for the compared agent-based methods, their internal token usage and computational costs may differ. Moreover, the absolute success rates remain limited, indicating that reliable secure code generation and automated vulnerability repair remain open challenges. SecAwareCoder should therefore be regarded as a research framework rather than a production-ready defense, and its outputs still require independent testing, complementary security analysis, and human review.

\section{Conclusion}
We presented \textbf{CodeSecEval}, a curated benchmark for evaluating \emph{secure code generation} through precise, reproducible, and execution-based protocols. CodeSecEval comprises 255 Python tasks spanning 77 CWE categories and provides paired insecure and secure implementations together with executable test suites that separately assess functional correctness and vulnerability-targeted security behaviors. This design supports automated evaluation using Pass@k, F-Pass@1, and S-Pass@1, while also enabling the study of insecure-code repair.

Building on CodeSecEval, we proposed \textbf{SecAwareCoder}, an agent-based framework that incorporates task-adaptive threat modeling into code generation and refinement. SecAwareCoder derives task-grounded vulnerability hypotheses, uses them to guide constraint-aware generation and security-aware test synthesis, and leverages execution feedback to perform targeted repairs. Experiments across various LLM backbones show that SecAwareCoder consistently achieves the highest overall Pass@1 among the evaluated methods and generally improves performance on vulnerability-targeted security tests while maintaining competitive functional correctness.

\section{Data Availability}
CodeSecEval is publicly available on Hugging Face at \url{https://huggingface.co/datasets/JasonWang1/CodeSecEval}, while the implementation of SecAwareCoder is publicly available at \url{https://github.com/CY-SCUT-MS/CodeSecEval}.

\begin{acks}
This research is supported by the Science and Technology Planning Project of Guangdong Province (2025B0101120003), the Guangdong Provincial Fund for Basic and Applied Basic Research—Regional Joint Fund Project (Key Project) (2023B1515120078), the National Natural Science Foundation of China (62402185, 62476097), the Guangdong Provincial Natural Science Foundation for Outstanding Youth Team Project (2024B1515040010), and the Fundamental Research Funds for the Central Universities (2025ZYGXZR064). This research is also sponsored by the CIPS-SMP-Zhipu Large Model Fund.
\end{acks}

\bibliographystyle{ACM-Reference-Format}
\bibliography{sample-base}

\end{document}